\documentclass[10pt,superscriptaddress,twocolumn,amsmath,amssymb,aps,prl,showpacs]{revtex4-1}
\usepackage{mathrsfs}
\usepackage{graphicx}
\usepackage{dcolumn}
\usepackage{bm}

\newcommand{\be}{\begin{equation}}
\newcommand{\ee}{\end{equation}}
\newcommand{\bea}{\begin{eqnarray}}
\newcommand{\eea}{\end{eqnarray}}

\begin{document}
\title{Fluctuation Effects on the Transport Properties of Unitary Fermi Gases}
\author{Boyang Liu}
\affiliation{Institute for Advanced Study, Tsinghua University,
Beijing, 100084, China}

\author{Hui Zhai}
\affiliation{Institute for Advanced Study, Tsinghua University,
Beijing, 100084, China}

\author{Shizhong Zhang}
\affiliation{Department of Physics and Center of Theoretical and
Computational Physics, The University of Hong Kong, Hong Kong,
China}

\date{\today}
\begin{abstract}
In this letter, we investigate the fluctuation effects on the transport properties of unitary Fermi gases in the vicinity of the superfluid transition temperature $T_{\rm c}$. Based on the time-dependent Ginzburg-Landau formalism of the BEC-BCS crossover, we investigate both the residual resistivity below $T_{\rm c}$ induced by phase slips and the paraconductivity above $T_{\rm c}$ due to pair fluctuations. These two effects have been well studied in the weak coupling BCS superconductor, and here we generalize them to the unitary regime of ultracold Fermi gases. We find that while the residual resistivity below $T_{\rm c}$ increases as one approaches the unitary limit, consistent with recent experiments, the paraconductivity exhibits non-monotonic behavior. Our results can be verified with the recently developed transport apparatus using mesoscopic channels.
\end{abstract}
\maketitle

In the past decade, one of the most exciting topics in cold atom
physics is the unitary Fermi gas characterized by the absence of a
small perturbation parameter and strong pairing
fluctuations~\cite{Giorgini, Zwerger, Taylor}. Thermodynamic
properties of the unitary Fermi gases have been well
studied~\cite{Horikoshi2010,Nascimbene2010,Ku2012} and are shown to
be universal~\cite{Ho2004}. Several experiments have also started to
investigate the transport properties of the unitary Fermi gases,
including first and second sound~\cite{Sidorenkov2013}, shear
viscosity~\cite{Cao2011} and spin
diffusion~\cite{Sommer2011,Koschorreck2013,Bardon2014}. In the later
two cases, apparent lower quantum limits have been observed in
experiments. Recently, a mesoscopic channel between two bulk unitary
Fermi gases has been constructed and a drop of resistance below
superfluid transition temperature $T_{\rm c}$ has been
seen~\cite{Stadler2012}. With the same setup, contact
resistance~\cite{Brantut2012}, quantized
conductance~\cite{Krinner2014} and thermoelectric
effect~\cite{Brantut2013} have also been observed. These
experimental developments offer new opportunities to study
mesoscopic transport phenomena with the flexibility of cold atoms.

Historically, fluctuation effects on transport properties have been well studied in the weak coupling superconductors~\cite{Tinkham}. Two well-known examples in the vicinity of superconducting transition temperature $T_\text{c}$ are: (a) Below $T_\text{c}$, a finite resistance appears due to phase slip induced by thermal fluctuations, known as Langer-Ambegaokar-McCumber-Halperin (LAMH) effect~\cite{Langer, McCumber}; (b) Above $T_\text{c}$, conductivity is enhanced due to Cooper pair fluctuations, often called  ``paraconductivity" as was first studied by Aslamazov and Larkin~\cite{Aslamazov, Skocpol}. In this Letter, we extend the above calculations to the unitary regime and show how the enhanced pair fluctuation modifies the above two effects. Our main conclusions are:

(i) For the appearance of resistance below $T_\text{c}$, we find that in the unitary regime, the resistivity drops much slower than in the BCS limit as temperature decreases.

(ii) For the enhancement of conductivity above $T_\text{c}$, we find this paraconductivity changes {\em non-monotonically} from the BCS limit to the unitary regime, and a minimum exists in between.

\emph{Time-dependent Ginzburg-Landau Theory}. Our derivation is based on the time-dependent Ginzburg-Landau (TDGL) theory of BEC-BCS crossover~\cite{Sademelo}. The partition function of the unitary Fermi gas can be written as $\mathcal{Z}=\int D[\bar\psi_\sigma,\psi_\sigma]\exp(-S[\bar\psi_\sigma,\psi_\sigma]),$
where $S[\bar\psi_\sigma,\psi_\sigma]=\int d\tau d^3{\bf x}\{\bar\psi_\sigma(\partial_\tau-\frac{\nabla^2}{2m}-\mu)\psi_\sigma-g\bar\psi_\uparrow \bar\psi_\downarrow\psi_\downarrow\psi_\uparrow\}$. Here $\tau$ is the imaginary time and $\mu$ is chemical potential. As usual, $g$ is related to $s$-wave scattering length $a_\text{s}$ by $1/g=-m/(4\pi a_\text{s})+\sum_{{\bf k}}1/(2\epsilon_{\bf k})$ with $\epsilon_{\bf k}={\bf k}^2/2m$. Introducing Hubbard-Stratonovich fields $\Delta(\tau, {\bf x})$ to decouple the interaction term in the Cooper channel and then integrating out the fermions, we obtain an effective theory for the bosonic field $\Delta(\tau, {\bf x})$ representing the bosonic Cooper pair field. In the vicinity of $T_\text{c}$ where $\Delta$ is small, we can expand the action in powers of $\Delta$, as well as its spatial and time derivatives (after Wick rotation),
\be
S[\bar\Delta,\Delta]=\int dt d^3{\bf x}\big\{\bar\Delta(\gamma\partial_t-\frac{\nabla^2}{2m^\ast}-r)\Delta+\frac{b}{2}\bar\Delta \bar\Delta\Delta\Delta\big\},
\label{eq:GL}
\ee
where $\gamma=\gamma_1+i\gamma_2$ is complex in general. All the parameters $\gamma$, $m^*$, $r$ and $b$ can be expressed in terms of $\mu$, $T$ and $\zeta\equiv 1/(k_{\rm F}a_\text{s})$~\cite{supp}. In the following, we will focus in the vicinity of the superfluid transition temperature, $T\approx T_{\rm c}$ and as a result, $\mu(T)\approx \mu(T_{\rm c})$. We determine both $T_{\rm c}$ and $\mu(T_{\rm c})$ within the Nozi\`{e}res-Schmitt-Rink (NSR) scheme~\cite{NSR}.

\begin{figure}[t]
  \includegraphics[width=7cm]{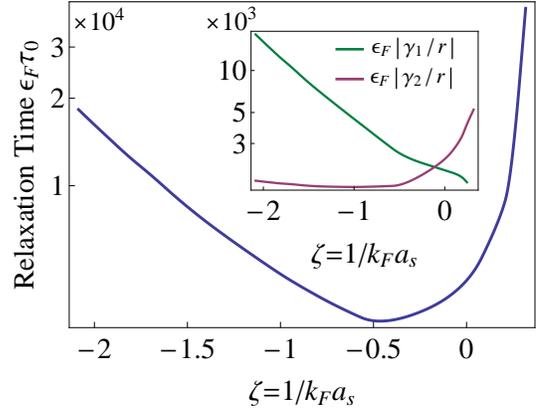}
  \caption{(Color online) Relaxation time $\tau_0$ as a function of $\zeta\equiv 1/(k_\text{F}a_\text{s})$, in unit of $1/\epsilon_{\rm F}$. Inset: the parameters $\gamma_1/r$ and $\gamma_2/r$ as a function of $\zeta$.  The temperature of the system is fixed at $1-T/T_{\rm c}=10^{-3}$.}
 \label{fig:relaxation}
\end{figure}

The real part $\gamma_1$ describes the damping of Cooper pairs due to coupling to fermionic quasi-particles. It can be shown that $\gamma_1$ is proportional to $\sqrt \mu\Theta(\mu)$ \cite{Sademelo}, where $\Theta(\mu)$ is the Heaviside step function. As a result, around unitarity and in the BCS side where $\mu>0$, the Cooper pairs have finite life time; while in the BEC limit where $\mu<0$, $\gamma_1=0$ and the Cooper pairs (molecules) are infinitely long-lived within NSR. The imaginary part $\gamma_2$ represents a propagating behavior and is given by
\be
\gamma_2=-\mathcal {P}\int \frac{d^3{\bf k}}{(2\pi)^3}\frac{1-2N(\xi_{\bf k})}{4\xi_{\bf k}^2},
\ee
where ``$\mathcal {P}$" denotes principle value. $N(\xi_{\bf k})=(\exp(\beta\xi_{\bf k})+1)^{-1}$ is the Fermi distribution function and $\xi_{\bf k}=\epsilon_{\bf k}-\mu$. In the BCS limit $\zeta\to-\infty$, $\mu\gg\Delta$, the integrand is roughly antisymmetric with respect to the Fermi surface $\epsilon_{\bf k}=\mu$, a manifestation of particle-hole symmetry of the BCS state. Consequently, $\gamma_2\simeq0$. As $\zeta$ increases towards the unitarity and the BEC side, $\gamma_2$ gradually increases from zero, due to increasing violation of particle-hole symmetry. The behaviors of $\gamma_1$ and $\gamma_2$ as a function of $\zeta$ are shown in the inset of Fig. \ref{fig:relaxation}.

\textit{Relaxation time.} As will be shown later, the relaxation time of the pairing field $\Delta(t,{\bf x})$ plays an important role in both the LAMH effect and paraconductivity. In the following, we derive an expression for the relaxation time that is valid close to unitarity. As is known~\cite{Tinkham}, to maintain a non-zero thermal average of the pairing fluctuation, it is necessary to introduce the so-called Langevin force $\eta(t,{\bf x})$ into the TDGL equation,
\be
-\gamma\frac{\partial }{\partial t}\Delta=-\frac{\nabla^2}{2m^\ast}\Delta -r\Delta+b|\Delta|^2\Delta+\eta(t,{\bf x}),
\ee
where the Langevin force represents the driving force of the environment and is characterized by the ``white noise'' correlations~\cite{Larkin,Tinkham}
\be
\left\langle \eta^\ast(t,{\bf x})\eta(t^\prime,{\bf x}^\prime)\right\rangle=2\gamma_1k_BT\delta(t-t^\prime)\delta^3({\bf x}-{\bf x}^\prime).
\ee
With a straightforward calculation, one finds the correlation function for the order parameter~\cite{supp}
\begin{align}
\langle\bar\Delta_{\bf k}(t)\Delta_{\bf k}(0)\rangle=\frac{k_BT}{k^2/2m+|r|}\exp\left[-\Big(\frac{1}{\tau_k}+\frac{1}{i\tau_k^\prime}\Big)t\right],
\label{eq:correlator}
\end{align}
where
\be
\tau_k=\frac{\gamma_1^2+\gamma_2^2}{\gamma_1(k^2/2m+|r|)}, ~~~\tau_k^\prime=\frac{\gamma_1^2+\gamma_2^2}{\gamma_2(k^2/2m+|r|)}.
\ee
$\tau_k$ represents the temporal decay of the ${\bf k}$-th Fourier component of the order parameter, while $\tau_k^\prime$ characterizes its propagating behavior. In the limit ${\bf k}\rightarrow 0$, we obtain
\be
\tau_0=\frac{\gamma_1^2+\gamma_2^2}{\gamma_1|r|}.
\label{eq:rt}
\ee

In the BCS limit $\zeta\to-\infty$, $\gamma_2\approx 0$ and the relaxation time $\tau_0$ only depends on $\gamma_1$ and can be reduced to $\tau_{\rm BCS}=\gamma_1/r$. Furthermore, in the same limit, $\gamma_1\approx m\beta k_\text{F}/(16\pi)$, $r\approx m k_\text{F}(T_\text{c}-T)/(2\pi^2T_\text{c})$~\cite{supp}, as a result, $\tau_{\rm BCS}={\pi}/[8k_B(T_c-T)]$, consistent with the weak coupling results~\cite{McCumber}. Away from the BCS limit, $\tau_0$ depends on both $\gamma_1$ and $\gamma_2$. As shown in Fig. \ref{fig:relaxation}, as $\zeta$ increases from the BCS limit toward the unitary regime, $\tau_0$ first decrease as $\gamma_1/|r|$, and then increases as $\gamma^2_2/(\gamma_1|r|)$. A minimum of $\tau_0$ occurs between the BCS limit and the unitary regime when $\gamma_1\approx \gamma_2$. In the BEC side when $\mu<0$, $\tau_k\to\infty$, indicating an undamped bosonic mode. To capture the effect of damping, it is necessary to go beyond the NSR scheme, which we will not attempt here. Rather we focus around unitarity, where our calculation applies.

\emph{Residual resistance below $T_c$}. To simplify our investigation, let us consider the residual resistance of a quasi-one-dimensional unitary Fermi gases of cross-section area $A$ and linear dimension $L$. The residual resistance below $T_{\rm c}$ is due to the thermally activated phase slips. The net effect of these events is to lower the current of the state, and as a result, a voltage drop must be sustained in order to maintain a steady current~\cite{Tinkham}. In another words, a finite resistance appears below $T_c$. Such a theory is developed by LAMH and later confirmed by experiments on BCS superconductors~\cite{Newbower}.

Within LAMH theory, residual resistivity due to the phase slips is given by \cite{Tinkham,Langer, McCumber,supp}
\be
\rho(T\lesssim T_\text{c})=\frac{2\pi A\Omega}{L k_BT}\exp\left[-\frac{\Delta F_0}{k_BT}\right].
\label{eq:resistivity}
\ee
$\Delta F_0$ is the lowest free-energy barrier to create one phase slip. Its analytic expression was derived by Langer and Ambegaokar: $\Delta F_0=\frac{8\sqrt2}{3}\frac{r^2}{2b}A \xi,$ where $r^2/2b$ is the condensation energy density and $\xi=1/\sqrt{2m^\ast|r|}$ is the Ginzburg-Landau coherence length. $\Delta F_0$ is roughly the condensation energy in a volume $A\xi$. $\Omega$ is the so-called ``attempt frequency",  originally derived by McCumber and Halperin \cite {McCumber} as $\Omega=\frac{L}{\xi}\sqrt{\frac{\Delta F_0}{k_B T}}\frac{1}{\tau_{\rm BCS}}$. In our case the relaxation time $\tau_{\rm BCS}$ has to be replaced by $\tau_0$ derived above.

\begin{figure}[t]
  \includegraphics[width=7cm]{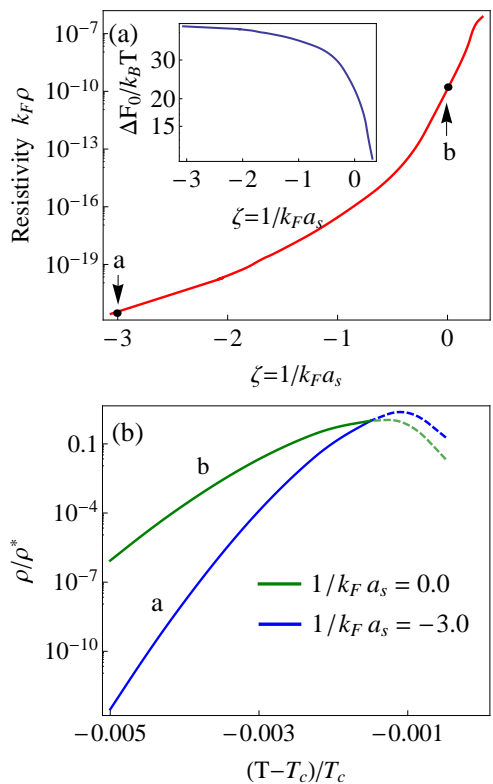}
  \caption{(Color online) (a) Resistivity (and in the inset $\Delta F_0/k_BT$) as a function of the coupling $1/k_\text{F}a_\text{s}$ with the temperature below $T_\text{c}$ by
  $(T_\text{c}-T)/T_\text{c}=5\times10^{-3}$. (b) Resistivity as a function of temperature. The curves ``a" and ``b" correspond to two different scattering lengths in marked in (a). Here we take $k^2_\text{F}A=10^6$.
  }
 \label{fig:resistivity}
\end{figure}
\begin{figure}[t]
  \includegraphics[width=7cm]{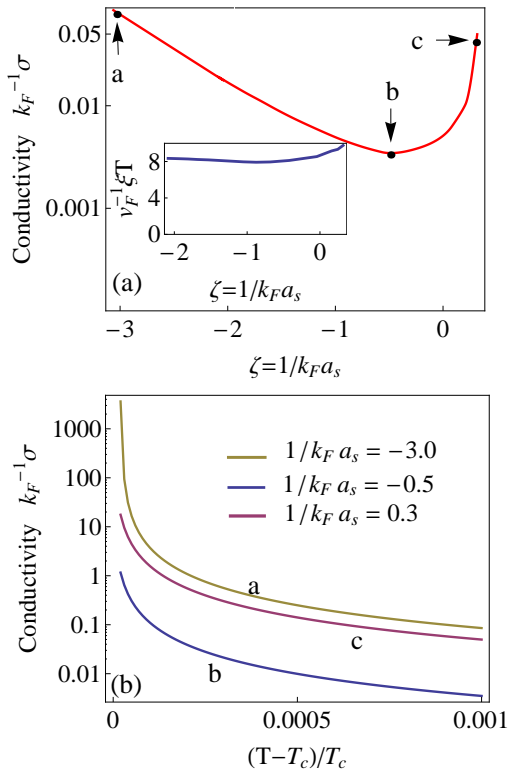}
  \caption{(Color online) (a) The fluctuation induced paraconductivity as a function of the coupling $1/(k_\text{F}a_\text{s})$ with the temperature above $T_\text{c}$ by
 $(T-T_\text{c})/T_\text{c}=10^{-3}$. (b) The fluctuation induced paraconductivity as a function of temperature. Curves marked by ``a", ``b" and ``c" correspond to three different scattering lengths marked in (a). Here we take $k^2_\text{F}A=10^6$.}
 \label{fig:conductivity}
 \end{figure}

Let us first investigate the dependence of $\rho$ on the interaction parameter $\zeta$. To do that, we first fix the temperature below $T_{\rm c}$ by $1-T/T_{\rm c}=5\times 10^{-3}$. Then it is clear from Eq.~\ref{eq:resistivity}, the resistivity depends on several ratios $\Delta F_0/k_B T$, $v_{\rm F}^{-1}T\xi$ and $\epsilon_{\rm F}\tau_0$. Within our calculation, $\epsilon_{\rm F}\tau_0$ changes only by a factor of $3\sim 4$ from BCS to the unitary regime. On the other hand, if one uses the weak coupling expression for $\xi=v_{\rm F}/\pi\Delta$ and the fact that $T\approx T_{\rm c}\sim \Delta$ the parameter $v_{\rm F}^{-1}T\xi$ then remains almost a constant. Numerical calculation shows that in the regime of $\zeta$ considered, $v_{\rm F}^{-1}T\xi$ changes only a few percent; see inset of Fig.\ref{fig:conductivity} (a). Now, the most important dependence is on $\Delta F_0/k_B T$, since it appears on the exponential factor. Detailed calculation shows that $\Delta F_0/k_B T$ changes by a factor about $3\sim 4$ in the relevant regime; see inset of Fig.\ref{fig:resistivity}(a). Taking into account all these dependences, we find that from the BCS side to unitarity, the fluctuation induced residual resistivity increases rapidly by several orders of magnitude, as shown in Fig.\ref{fig:resistivity}.

Now let us look at the temperature dependences of the residual resistivity. In Fig.\ref{fig:resistivity} (b), we plot the resistivity $\rho$ in the BCS limit ($\zeta=-3$) and at unitarity ($\zeta=0$), normalized to their respective values $\rho^*$ at $1-T/T_{\rm c}=1.5\times 10^{-3}$. We observe that as temperature decreases, the resistivity drops much slower at unitarity than in the BCS limit. We also note that in Fig. \ref{fig:resistivity} (b), there is a unusual drop of resistivity (marked by the dashed lines) when temperature is very close to $T_\text{c}$. This is because the LAMH theory fails very close to $T_c$ \cite{Skocpol}.

The above two observations at unitarity, the increased residual resistivity and its slower decrease as a function of temperature, suggest the more pronounced role of superconducting fluctuations below $T_{\rm c}$ at unitarity in comparison with the BCS limit. The increase of resistivity is monotonic as one approaches unitarity from the BCS side, in accordance with our general expectations.  In fact, as was discovered recently, close to unitarity when $A/\xi^2\gg 1$, the energetically more favorable defects is a solitonic vortex~\cite{Yefsah2014,Ku2014,Bulgac2014}, instead of the phase soliton in the BCS regime where $A/\xi^2\lesssim 1$. Thus our estimation of $\Delta F_0$ is an overestimate of the defect energy and the residual resistivity should in fact increase more rapidly close to unitarity and decrease even slower as the temperature is lowered. However, when we turn to the fluctuation induced conductivity above $T_{\rm c}$, as we shall show shortly, the effect is not monotonic and in fact exhibits a minimum in between.

\emph{Enhanced Conductivity above $T_c$}. Above $T_\text{c}$, in addition to the usual conductivity given by normal fermions, there will be an extra contribution to conductivity due to thermal fluctuation of Cooper pairs field $\Delta(x, t)$, known as paraconductivity. We introduce the fluctuating supercurrent $J(t)$ along one of spatial direction, say, $\hat{x}$, where $J_x(t)$ is given by $J_x(t)=\frac{1}{m^\ast}\sum_{\bf k}k_x|\Delta_{\bf k}(t)|^2$. The fluctuation induced paraconductivity can be directly calculated using the Kubo formalism as
\be
\sigma_{xx}(\omega)=\frac{1}{k_BT}\int^\infty_0dt\langle J_x(t)J_x(0)\rangle\cos(\omega t).
\ee
A straightforward calculation yields the current-current correlation function as  \cite{Tinkham,Larkin}
\be
\langle J_x(t)J_x(0)\rangle=\left(\frac{1}{m^\ast}\right)^2\sum_{\bf k}k_x^2|\langle \bar\Delta_{\bf k}(t)\Delta_{\bf k}(0)\rangle|^2.
\label{Jcorrelation}
\ee
While $\langle\Delta\rangle=0$ for $T>T_{\rm c}$, the thermal fluctuation of Cooper pair field $\Delta(t,{\bf x})$ renders a non-zero value of the time-correlation function $\langle \bar\Delta_{\bf k}(t)\Delta_{\bf k}(0)\rangle$, as found previously in Eq. (\ref{eq:correlator}). This yields a non-zero contribution to the conductivity above $T_{\rm c}$,
\be
\sigma_{xx}(\omega)=\frac{1}{k_BT{m^\ast}^2}\sum_{\bf k}\left[\frac{k_x k_BT}{k^2/2m^\ast+|r|}\right]^2\frac{\tau_k/2}{1+(\tau_k\omega/2)^2}.
\label{sigma}
\ee
We note that only $\tau_k$, which characterize the temporary decay of the order parameter correlation function, contributes to the conductivity. Specializing to the quasi-one-dimensional and considering the DC component $\sigma_0\equiv \sigma(\omega=0)$, the paraconductivity can be written as
\begin{align}
\sigma_0=2k_BT\tau_0\int^{+\infty}_{-\infty}\frac{dk_x}{2\pi}\frac{k_x^2\xi^4}{A(k^2\xi^2+1)^3}=\frac{k_BT\xi}{8A}\tau_0,
\label{eq:paraconductivity}
\end{align}
with $\tau_0$ given by Eq.(\ref{eq:rt}). To see how $\sigma_0$ changes as a function of interaction strength $\zeta$, let us fix the temperature slightly above $T_{\rm c}$, $1-T/T_{\rm c}=10^{-3}$. As we show in Fig. \ref{fig:conductivity} (a), as one goes from the BCS limit to the unitary regime, fluctuation induced paraconductivity first decreases and then increases, in comparison with the monotonic behavior of phase slip induced resistivity below $T_{\rm c}$. The similar dependence on $\zeta$ of $\sigma_0$ and the relaxation time $\tau_0$ can be understood in the following way. According to Eq.\ref{eq:paraconductivity}, $\sigma_0$ is proportional to $\tau_0$ with the coefficient $T\xi$. Now, as we have shown before, since again $T\sim T_{\rm c}\sim \Delta$, $T\xi$ remains approximately a constant and as a result, $\sigma_0$ exhibits qualitatively the same dependence on $\zeta$ as $\tau_0$.

Now, let us look at the temperature dependences of $\sigma_0$ for various values of $\zeta$. In Fig. \ref{fig:conductivity} (b), we show $\sigma_0$ at three interaction strengths: $\zeta=-3$ (marked by a), $\zeta=-0.5$ (marked by b) which is at the minimal of $\sigma_0$ and $\zeta=0.3$ (marked by c). They all show rapid increase as one approaches $T_{\rm c}$ from above.

\emph{Discussions.} In this work we have discussed fluctuation effects on the transport properties of the unitary Fermi gas based on TDGL theory. At present, our results cannot be directly applied to the BEC limit since we have not taken properly into account the interactions between molecules. This leads to the infinite lifetime in the BEC side of the crossover where $\mu<0$. Furthermore, in the BEC limit, it is also important to take into account the correction to chemical potential arising from the molecular interaction. We shall leave this to a future investigation.

In recent ETH experiment they have observed drop of the resistance for unitary Fermi gas below $T_\text{c}$, but the drop is much slower compared with typical BCS superconductor~\cite{Stadler2012}, consistent with our findings, although their experimental situation is much complicated than what is discussed here. Namely, the finite resistance observed below $T_{\rm c}$ is due to the thermally activated phase slips which becomes much easier when close to unitarity, reflecting its enhanced superconducting fluctuations. Furthermore, we find that fluctuation induced conductivity (paraconductivity) above $T_\text{c}$ exhibits non-monotonic behavior as one approaches unitarity from the BCS side. This can be verified in the same ETH experimental setup.

\emph{Acknowledgements.} BY and HZ are supported by Tsinghua University Initiative Scientific Research Program, NSFC Grant No. 11174176, and NKBRSFC under Grant No. 2011CB921500. SZ is supported by the startup fund from the University of Hong Kong and collaborative research fund HKUST3/CRF/13G. HZ would like to thank Hong Kong University for hospitality where part of this work is finished.

\pagebreak

\begin{widetext}
\setcounter{equation}{0} \setcounter{figure}{0}
\setcounter{table}{0} \setcounter{page}{1} \makeatletter

\section{Supplementary Materials}

\subsection{Time-dependent Ginzburg-Landau theory of BEC-BCS
crossover} A time-dependent Ginzburg-Landau theory can be
constructed for the entire BEC-BCS crossover close to $T_{\rm
c}$~\cite{SSademelo}. The partition function takes the form
$\mathcal Z=\int
D[\bar\psi_\sigma,\psi_\sigma]e^{-S[\bar\psi_\sigma,\psi_\sigma]}$,
with \be S[\bar\psi_\sigma,\psi_\sigma]=\int d\tau d^3{\bf x}\Big
\{\bar\psi_\sigma(\partial_\tau-\frac{\nabla^2}{2m}-\mu)\psi_\sigma-g\bar\psi_\uparrow
\bar\psi_\downarrow\psi_\downarrow\psi_\uparrow\Big \}, \ee where
$\psi_\sigma$ are Grassman fields and $g$ is the contact interaction
between fermions of opposite spins. $\mu$ is the chemical potential
which is determined by requiring the number density to be equal to
$n$. To investigate the fluctuation effects in the Cooper channel,
we use Hubbard-Stratonovich transformation to decouple the
interaction term in the Cooper channel and then integrating out the
fermions. We obtain an effective theory for the bosonic field
$\Delta(\tau, {\bf x})$, which represents the cooper pair field.
Straightforward calculations yield the partition function of field
$\Delta$ as \be \mathcal Z=\int D(\bar\Delta,\Delta)\exp
\Big[-\frac{1}{g}\int d\tau d{\bf x}|\Delta|^2+\ln\det\hat
G^{-1}\Big], \ee where \be \hat{G}^{-1}=
\left(\begin{array}{cc}-\partial_\tau+\frac{\nabla^2}{2m}+\mu &
\Delta\\ \bar \Delta &
-\partial_\tau-\frac{\nabla^2}{2m}-\mu\end{array}\right) \ee is the
Gor'kov Green function.

In the vicinity of the phase transition the gap parameter $\Delta$
is small and an expansion in terms of $\Delta$ becomes possible.
Including both the spatial and time derivatives (after Wick
rotation) and retaining the parameter $\Delta$ up to the forth order
we obtain an effective action as \be S[\bar\Delta,\Delta]=\int dt
d^3{\bf
x}\Big\{\bar\Delta\big[\gamma\partial_t-\frac{\nabla^2}{2m^\ast}-r\big]\Delta+\frac{b}{2}\bar\Delta
\bar\Delta\Delta\Delta\Big\}, \label{eq:GL} \ee where
$\gamma=\gamma_1+i\gamma_2$ and all the parameters can be expressed
in terms of microscopic parameters as
\begin{align}
\gamma_1 &=\frac{m^{3/2}}{8\sqrt2\pi}\beta\sqrt \mu\Theta(2\mu),\\
\gamma_2 &=-\mathcal {P}\int \frac{d^3{\bf k}}{(2\pi)^3}\frac{1-2N(\xi_{\bf k})}{4\xi_{\bf k}^2},\\
\frac{1}{2m^\ast} &=\frac{1}{2m}\int\frac{d^3{\bf
k}}{(2\pi)^3}\Bigg\{\frac{1-2N(\xi_{{\bf k}})}{8\xi_{{\bf k}}^2}
+\frac{\frac{\partial N(\xi_{{\bf k}})}{\partial\xi_{\bf k}}}{4\xi_{\bf k}}+\frac{\frac{\partial^2N(\xi_{{\bf k}})}{\partial \xi^2_{\bf k}}\cdot\frac{{\bf k}^2}{2m}}{6\xi_{\bf k}}\Bigg\},\\
 r &=\frac{m}{4\pi a}+\int \frac{d^3{\bf k}}{(2\pi)^3}\Bigg\{\frac{1-2N(\xi_{\bf k})}{2\xi_{\bf k}}-\frac{1}{2\epsilon_{\bf k}}\Bigg\},\\
b &=\int\frac{d^3{\bf k}}{(2\pi)^3}\Bigg\{\frac{1-2N(\xi_{\bf
k})}{4\xi_{\bf k}^3}+\frac{\beta N(\xi_{\bf k})[N(\xi_{\bf
k})-1]}{2\xi_{\bf k}^2}\Bigg\}. \label{eq:GLparameter}
\end{align}
In above equation $N(\xi_{\bf k})=1/(\exp(\beta\xi_{\bf k})+1)$ is
the Fermi distribution function and $\xi_{\bf k}=\epsilon_{\bf
k}-\mu$ with $\epsilon_{\bf k}={\bf k}^2/2m$. Function
$\Theta(2\mu)$ is the heaviside step function. Notation ``$\mathcal
{P}$" in equation of $\gamma_2$ denotes the principle value.
Explicitly, the parameter $b$ is the result of loop calculation with
four fermion propagators \be
b=-\frac{1}{\beta^2}\sum_{\omega_n}\int\frac{d^3{\bf
k}}{(2\pi)^3}\frac{1}{(-i\omega_n+k^2/2m-\mu)^2}\frac{1}{(i\omega_n+k^2/2m-\mu)^2}.
\ee The other parameters $\gamma_1$, $\gamma_2$, $\frac{1}{2m^\ast}$
and $r$ are all derived from the inverse vertex function
$\Gamma^{-1}(\omega_n,{\bf k})$, which after the standard
renormalization which replace $g$ with the two-body scattering
length $a_{\rm s}$, is given by \be \Gamma^{-1}(\omega_n,{\bf
k})=-\frac{m}{4\pi a_{\rm s}}-\int \frac{d^3{\bf
k}}{(2\pi)^3}\left\{\frac{1-N(\epsilon_{\bf k}-\mu)-N(\epsilon_{\bf
k-q}-\mu)}{-i\omega_n+\epsilon_{\bf k}+\epsilon_{\bf
k-q}-2\mu}-\frac{1}{2\epsilon_{\bf k}}\right\}. \ee To derive the
time-dependent Ginzburg-Landau equation, we first analytically
continue vertex function to real frequency
$i\omega_n\to\omega+i0^+$. This procedure generates a time-dependent
term with parameter $\gamma=\gamma_1+i\gamma_2$. The $\gamma_2$ term
exhibits a propagating behavior. As long as $\mu>0$, $\gamma_1$ is
nonzero, which indicates a finite lifetime of the Cooper pairs.

Three important quantities that characterize the time-dependent
Ginzburg-Landau theory are relaxation time, coherence length and
condensation energy. The variation of the relaxation time as a
function of $\zeta\equiv 1/k_F a_s$ is illustrated in Fig. 1 of the
main text. Here we plot the variations of the coherence length and
condensation energy with respect to $1/k_F a_s$.
\begin{figure}[h]
\begin{center}
  \includegraphics[width=12cm]{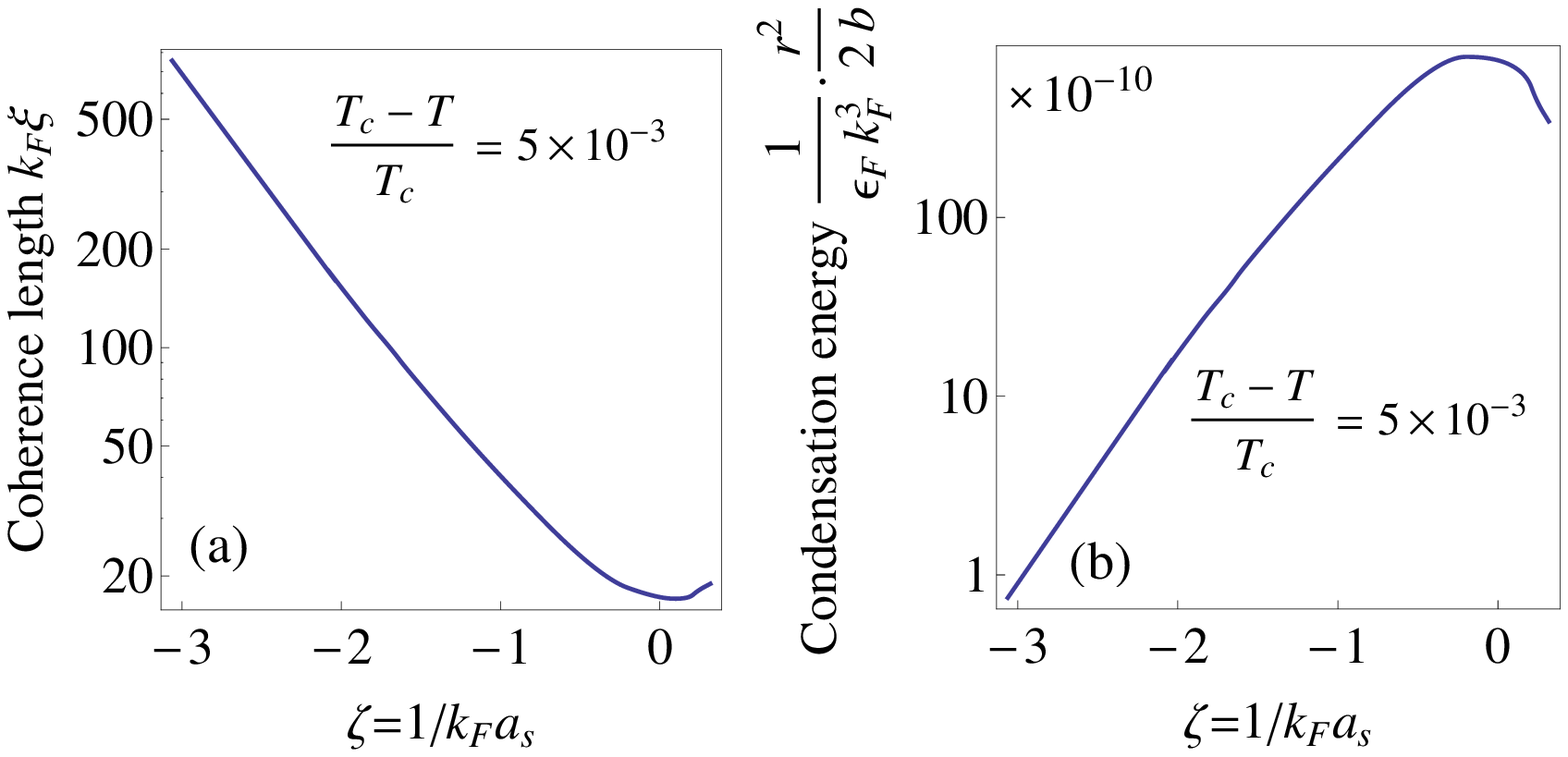}
  \caption{The coherence length and condensation energy as functions of $1/k_Fa_s$.}
  \label{fig:clce}
  \end{center}
 \end{figure}
In BCS and BEC limits all the parameters can be analytically derived
as shown in Table I.
\begin{table}[h]
\begin{center}
\begin{tabular}{|c|c|c|}\hline
\textbf{Parameters} & \textbf{BCS limit} & \textbf{BEC limit}\\
\hline
$\gamma_1$ &$ \nu(\epsilon_F)\cdot\frac{\pi}{8k_BT}$&$0$\\
\hline $\gamma_2$ & 0 &
$-\frac{\pi\nu(\epsilon_F)}{8\sqrt{\epsilon_F|\mu|}}$
\\ \hline $\frac{1}{2m^\ast}$ & $\frac{1}{2m}\cdot\frac{7\nu(\epsilon_F)\epsilon_F}{12\pi^2(k_BT)^2}\zeta(3)$ &
$\frac{1}{2m}\cdot\frac{\pi\nu(\epsilon_F)}{16\sqrt{\epsilon_F|\mu|}}$
\\ \hline $r$ & $\nu(\epsilon_F)\ln\frac{T_c}{T}$ & $\frac{\pi\nu(\epsilon_F)}{2\sqrt2\sqrt{\epsilon_F}}\cdot(\frac{1}{\sqrt m a_s}-\sqrt{2|\mu|})$\\\hline
$b$ & $\frac{7\nu(\epsilon_F)}{8\pi^2(k_BT)^2}\cdot\zeta(3)$
&$\frac{\pi\nu(\epsilon_F)}{32\sqrt{\epsilon_F}|\mu|^{3/2}}$\\
\hline
\end{tabular}
\end{center}
\caption{Asymptotic behaviors of the parameters in the
time-dependent Ginzburg-Landau theory in the BCS and BEC limits.}
\end{table}

Using the asymptotic expressions above we can derive a
Gross-Pitaevski equation from Eq. (\ref{eq:GL}) at BEC limit. By
defining $\Psi=\sqrt{|\gamma_2|}\Delta$ we obtain \bea
-i\partial_t\Psi-\frac{\nabla^2}{2M}\Psi-\tilde\mu\Psi+U|\Psi|^2\Psi=0.\eea
The parameters in above equation can be calculated as \bea &&
\frac{1}{2M}=\frac{1}{2m^\ast|\gamma_2|}=\frac{1}{2m}\cdot\frac{\pi\nu(\epsilon_F)}{16\sqrt{\epsilon_F|\mu|}}\cdot
\frac{8\sqrt{\epsilon_F|\mu|}}{\pi\nu(\epsilon_F)}=\frac{1}{4m},\cr
&&\tilde\mu=\frac{r}{|\gamma_2|}=\frac{\pi\nu(\epsilon_F)}{2\sqrt2\sqrt{\epsilon_F}}\cdot(\frac{1}{\sqrt
m a_s}-\sqrt{2|\mu|})\cdot
\frac{8\sqrt{\epsilon_F|\mu|}}{\pi\nu(\epsilon_F)}\approx 0, \cr&&
U=\frac{b}{\gamma_2^2}=\frac{\pi\nu(\epsilon_F)}{32\sqrt{\epsilon_F}|\mu|^{3/2}}\cdot
\Big(\frac{8\sqrt{\epsilon_F|\mu|}}{\pi\nu(\epsilon_F)}\Big)^2=\frac{4\pi
a_b}{M},\eea where the boson mass $M=2m$, the binding energy
$E_b=1/(ma_s^2)$ and the boson scattering length $a_b=2a_s$.

\subsection{Langer-Ambegaokar-McCumber-Halperin theory} In this part
we give a simple derivation of the
Langer-Ambegaokar-McCumber-Halperin theory~\cite{STinkham}. We start
our analysis from the static Ginzburg-Landau free energy functional
\bea F=\int d^3{\bf
x}\Big\{\bar\Delta\big[-\frac{\nabla^2}{2m^\ast}-r\big]\Delta+\frac{b}{2}\bar\Delta
\bar\Delta\Delta\Delta\Big\}.\eea Minimizing the free energy with
respect to the field $\Delta$ yields the time-independent
Ginzburg-Landau equation $-\frac{\nabla^2}{2m^\ast}\Delta-r\Delta+b
|\Delta|^2\Delta$. For a neutral system the current density is
written as $J=\frac{1}{2m^\ast
i}[\bar\Delta\nabla\Delta-\Delta\nabla\bar\Delta]$. A uniform
constant-current solution of the 1D Ginzburg-Landau equation can be
written as $\Delta_k=f_k\exp(ikx)$ with $f_k^2=(r-k^2/2m^\ast)/b$,
where $k$ is the allowed wave vector along the $x$ direction with
periodic boundary condition $\Delta(0)=\Delta(L)$. $L$ is the length
of the 1D tunnel. The current density subject to the solution of
$\Delta_k=f_k\exp(ikx)$ is $J=\frac{k(r-k^2/2m^\ast)}{m^\ast b}$.
This current has a maximum value $J_c=(2r/3)^{3/2}/\sqrt mb$ at
$k_c=\sqrt{2mr/3}$.

For $J<J_c$, the stead state is that of a persistent current without
any dissipation. According to the Josephson relation, this
corresponds to a definite phase twisting between the two ends of the
superconducting wire. Close to $T_{\rm c}$, thermal fluctuations can
either add or remove one more twist by $2\pi$ in the wire, with an
free energy barrier $\Delta F_0$ that is determined by Langer and
Ambegaokar~\cite{SLanger}. In the presence of supercurrent, there is
now a difference between the free energy barriers between the adding
($\Delta F_+$) or removing ($\Delta F_-$) of an extra $2\pi$-twist,
which is given by \be \delta F\equiv \Delta F_+-\Delta F_-=2\pi A J,
\ee where $A$ is the cross-section area of the 1D channel. The
prefactor $\Omega$ is the attempt frequency as discussed in the main
text. As a result, the rate of phase decreasing by $2\pi$ is
slightly larger than the one of phase increasing by $2\pi$. The
changing of phase difference between two ends per unit time is
\begin{align}
\frac{d\phi_{12}}{dt} &=\Omega \left[\exp\left(-\frac{\Delta
F_0}{k_BT}+\frac{\delta F}{2k_BT}\right)-\exp\left(-\frac{\Delta
F_0}{k_BT}-\frac{\delta F}{2k_BT}\right)\right]\\
&=2\Omega\sinh\left(\frac{\delta
F}{2k_BT}\right)\exp\left[-\frac{\Delta F_0}{k_BT}\right],
\end{align}
where $\phi_{12}$ is the difference between the order parameter
phases at the two ends of the channel. A steady current state can be
achieved when a small chemical potential difference $\Delta\mu$ is
applied on the ends of the wire. The resistivity can then be defined
as \be \rho\equiv\frac{\Delta \mu}{L J}
=\frac{2\Omega}{LJ}\sinh\left(\frac{\delta
F}{2k_BT}\right)\exp\left[-\frac{\Delta F_0}{k_BT}\right]. \ee For
small current the resistivity can be approximated as \bea
\rho\simeq\frac{2\pi\Omega A}{L k_BT}\exp\left[-\frac{\Delta
F_0}{k_BT}\right]\label{eq:rho}.\label{eq:resistivity}\eea

\end{widetext}
\end{document}